\documentclass[aps,pra,twocolumn,superscriptaddress]{revtex4-2}
\usepackage{graphicx}
\usepackage{amsmath}
\usepackage{amsbsy}
\usepackage{array}
\usepackage{amssymb}
\usepackage{multirow}
\usepackage{verbatim}
\usepackage{caption} 
\usepackage{subcaption} 
\usepackage{float} 
\usepackage{hyperref}
\usepackage{epstopdf}

\hypersetup{
	colorlinks=true,  
	linkcolor=blue,   
	citecolor=blue,   
	urlcolor=blue,    
	pdfborder={0 0 0}, 
}

\begin{document}
\title{Measuring network quantum steerability utilizing artificial neural networks}

\author{Mengyan Li}
\affiliation{School of Science, Beijing University of Posts and Telecommunications, Beijing 100876, China}
\affiliation{Key Laboratory of Mathematics and Information Networks, Beijing University of Posts and Telecommunications, Ministry of Education, Beijing 100876, China}
\affiliation{State Key Laboratory of Networking and Switching Technology, Beijing University of Posts and Telecommunications, Beijing, 100876, China}

\author{Yanning Jia}
\affiliation{School of Science, Beijing University of Posts and Telecommunications, Beijing 100876, China}
\affiliation{Key Laboratory of Mathematics and Information Networks, Beijing University of Posts and Telecommunications, Ministry of Education, Beijing 100876, China}
\affiliation{State Key Laboratory of Networking and Switching Technology, Beijing University of Posts and Telecommunications, Beijing, 100876, China}

\author{Fenzhuo Guo}\email{gfenzhuo@bupt.edu.cn}
\affiliation{School of Science, Beijing University of Posts and Telecommunications, Beijing 100876, China}
\affiliation{Key Laboratory of Mathematics and Information Networks, Beijing University of Posts and Telecommunications, Ministry of Education, Beijing 100876, China}
\affiliation{State Key Laboratory of Networking and Switching Technology, Beijing University of Posts and Telecommunications, Beijing, 100876, China}

\author{Haifeng Dong}
\affiliation{School of Instrumentation Science and Opto-Electronics Engineering, Beihang University, Beijing, 100191, China}

\author{Sujuan Qin}
\affiliation{State Key Laboratory of Networking and Switching Technology, Beijing University of Posts and Telecommunications, Beijing, 100876, China}

\author{Fei Gao}
\affiliation{State Key Laboratory of Networking and Switching Technology, Beijing University of Posts and Telecommunications, Beijing, 100876, China}

\begin{abstract}

	Network quantum steering plays a pivotal role in quantum information science, enabling robust certification of quantum correlations in scenarios with asymmetric trust assumptions among network parties. The intricate nature of quantum networks, however, poses significant challenges for the detection and quantification of steering. In this work, we develop a neural network-based method for measuring network quantum steerability, which can be generalized to arbitrary quantum networks and naturally applied to standard steering scenarios. Our method provides an effective framework for steerability analysis, demonstrating remarkable accuracy and efficiency in standard bipartite and multipartite steering scenarios. Numerical simulations involving isotropic states and noisy GHZ states yield results that are consistent with established findings in these respective scenarios. Furthermore, we demonstrate its utility in the bilocal network steering scenario, where an untrusted central party shares two-qubit isotropic states of different visibilities, $\nu$ and $\omega$, with trusted endpoint parties and performs a single Bell state measurement. Through explicit construction of a network local hidden state model derived from numerical results and incorporation of the entanglement properties of network assemblages, we analytically demonstrate that the network steering thresholds are determined by the curve $\nu \omega = {1}/{3}$ under the corresponding configuration.
	
\end{abstract}
\maketitle

\section{\label{sec1:level1}INTRODUCTION}

	\begin{figure*}[ht!]
		\centering
		\begin{subfigure}[b]{0.28\textwidth}
			\centering
			\includegraphics[width=\textwidth]{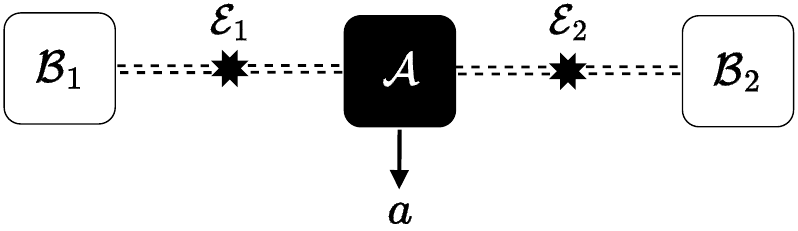}
			\caption{\centering}
			\label{fig1-(a)}
		\end{subfigure}
		\hspace{0.05cm}
		\begin{subfigure}[b]{0.16\textwidth}
			\centering
			\includegraphics[width=\textwidth]{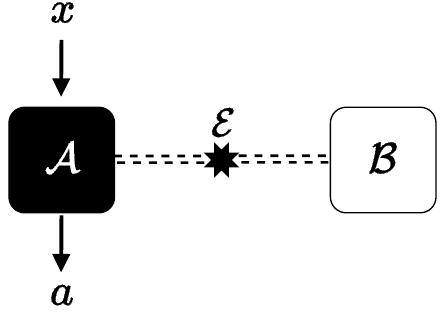}
			\caption{\centering}
			\label{fig1-(b)}
		\end{subfigure}
		\hspace{0.05cm}
		\begin{subfigure}[b]{0.16\textwidth}
			\centering
			\includegraphics[width=\textwidth]{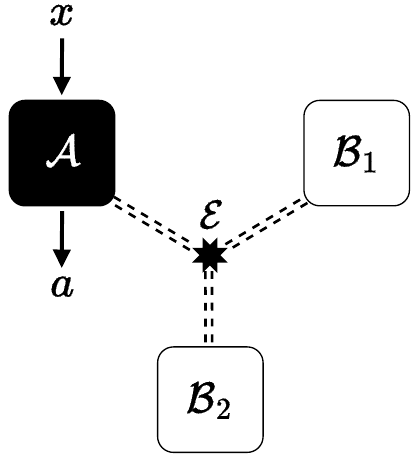}
			\caption{\centering}
			\label{fig1-(c)}
		\end{subfigure}
		\hspace{0.05cm}
		\begin{subfigure}[b]{0.16\textwidth}
			\centering
			\includegraphics[width=\textwidth]{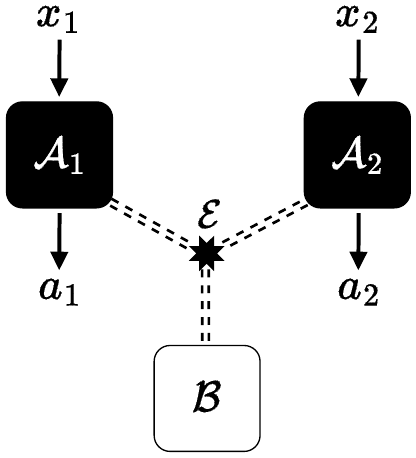}
			\caption{\centering}
			\label{fig1-(d)}
		\end{subfigure}
		\hspace{0.05cm}
		\caption{Diagrams of several steering scenarios. Black boxes denote untrusted parties, transparent boxes represent trusted parties, and all sources are assumed to be untrusted. For notational simplicity, subscripts are omitted when only a single trusted or untrusted party is involved. The scenarios depicted are: (a) Bilocal  network steering scenario with trusted endpoint parties and a central untrusted party. (b) Bipartite steering scenario. (c) Tripartite steering scenario with one untrusted party. (d) Tripartite steering scenario with two untrusted parties. }
		\label{fig1}
	\end{figure*}

	Quantum steering, an intermediate type of quantum correlation between quantum entanglement \cite{horodeckiQuantumEntanglement2009} and Bell nonlocality \cite{brunnerBellNonlocality2014}, was first introduced by Schrödinger \cite{Schrödinger_1935,Schrödinger_1936} and later rigorously formalized by Wiseman \textit{et al.}~\cite{wisemanSteeringEntanglementNonlocality2007}. Specifically, if the subnormalized states of the trusted party, following measurements performed by the untrusted party, do not admit a local hidden state (LHS) model, the system is said to exhibit steering \cite{wisemanSteeringEntanglementNonlocality2007}. This phenomenon holds potential applications in quantum information processing, such as quantum key distribution, sub-channel discrimination, quantum teleportation, and randomness certification \cite{cavalcantiQuantumSteeringReview2017,uolaQuantumSteering2020,xiangQuantumSteeringPractical2022}. In light of its importance, the detection and quantification of steering have garnered significant attention \cite{uolaQuantumSteering2020}. Early research primarily centered on single-source scenarios, commonly referred to as standard steering scenarios, leading to substantial theoretical progress. This includes inequality-based criteria  \cite{cavalcantiExperimentalCriteriaSteering2009,saundersExperimentalEPRsteeringUsing2010,cavalcanti2015analog}, operational measures like steering weight \cite{skrzypczykQuantifyingEinsteinPodolskyRosenSteering2014}, steering robustness \cite{pianiNecessarySufficientQuantum2015,cavalcantiQuantitativeRelationsMeasurement2016}, and steering cost \cite{dasCostEinsteinPodolskyRosenSteering2018}, as well as geometric quantification techniques \cite{jevticQuantumSteeringEllipsoids2014,jevticEinsteinPodolskyRosen2015,nguyenNecessarySufficientCondition2016,nguyenNonseparabilitySteerabilityTwoqubit2016,kuEinsteinPodolskyRosenSteeringIts2018,nguyenGeometryEinsteinpodolskyrosenCorrelations2019}.
	
	In 2021, Jones \textit{et al.}~\cite{jonesNetworkQuantumSteering2021} pioneered the investigation of quantum steering in network scenarios involving multiple independent sources, with a focus on linear networks comprising trusted endpoint parties and untrusted intermediate parties. Their work indicates that network steering manifests when the subnormalized states of the trusted parties, following measurements by all untrusted parties, do not admit a network local hidden state (NLHS) model. In subsequent years, researchers investigated network steering in two-forked tree-shaped networks \cite{yangQuantumSteeringTwoforked2023}, star networks \cite{chenDetectionNetworkGenuine2024,jiangQuantumSteeringStar2024}, swap-steering scenario \cite{sarkarNetworkQuantumSteering2024} and repeater networks \cite{liDetectingQuantumSteering2025}, thereby deriving several inequality-based criteria. Nevertheless, compared to standard steering scenarios, research on network steering scenarios remains insufficient. The inherent challenges stem from the intricate nature of network structures, particularly due to the constraints imposed by the independence of sources, which hinder the applicability of traditional methods. Fortunately, artificial neural networks (ANNs) \cite{Goodfellow-et-al-2016}, which are capable of effectively handling the complex relationships and high-dimensional parameter spaces characteristic of network steering scenarios, present a promising alternative for tackling these challenges.
	
	Hitherto, ANNs have been successfully applied to steering detection in standard steering scenarios~\cite{zhangDetectingSteerabilityBounds2022, haoImprovingSteerabilityDetection2023, wangDeepLearningHierarchy2024}. Specifically, Ref.~\cite{zhangDetectingSteerabilityBounds2022} utilized an error backpropagation neural network to construct high-performance quantum steering classifiers, enabling precise prediction of steering thresholds for generalized Werner states. Developing this further, Ref.~\cite{haoImprovingSteerabilityDetection2023} introduced an innovative aggregate class distribution neural network, enhancing the accuracy of quantum steering classification. Additionally, Wang \textit{et al.}~\cite{wangDeepLearningHierarchy2024} leveraged ANNs to investigate the impact of the number of measurement settings on the steerability of quantum states. 
	
	In this study, we develop a neural network-based method for measuring network quantum steerability. By introducing a well-defined metric \cite{kuEinsteinPodolskyRosenSteeringIts2018} to quantify the distance between pairs of quantum assemblages, we employ ANNs to search for the network assemblage that optimally approximates a given network assemblage while admitting the NLHS model. The distance between these two assemblages serves as a measure of network quantum steerability. This process is primarily accomplished by using ANNs to model causal information, as ANNs are inherently structured as directed acyclic graphs that naturally mirror the topology of quantum networks.
	
	To validate our method, we apply it to well-studied standard steering scenarios, including bipartite steering scenario involving isotropic states \cite{cavalcantiExperimentalCriteriaSteering2009,cavalcantiQuantitativeRelationsMeasurement2016,kuEinsteinPodolskyRosenSteeringIts2018} and multipartite steering scenarios involving noisy GHZ states \cite{cavalcantiDetectionEntanglementAsymmetric2015}. Numerical simulations yield results that demonstrate remarkable consistency with established findings in Refs.~\cite{cavalcantiExperimentalCriteriaSteering2009,cavalcantiQuantitativeRelationsMeasurement2016,kuEinsteinPodolskyRosenSteeringIts2018,cavalcantiDetectionEntanglementAsymmetric2015}, confirming both the accuracy and efficiency of our method. Furthermore, we demonstrate its utility in the bilocal network steering scenario, in which an untrusted central party shares two-qubit isotropic states of different visibilities with trusted endpoint parties and performs a single Bell state measurement (BSM). By explicitly constructing the NLHS model from numerical results and incorporating the entanglement properties of the network assemblages, we analytically determine the network steering thresholds in this configuration.

\section{\label{sec2:level1} GENERAL DEFINITION OF NETWORK QUANTUM STEERING SCENARIOS}
	
	Consider a network comprising $l$ sources, $m$ untrusted parties, and $n$ trusted parties. Each source $\mathcal{E}_i$ distributes the particles of a quantum state $\rho_i$ to its directly connected parties ($i = 1, 2, \dots, l$). Let each untrusted party be denoted as $\mathcal{A}_j$, who performs ${N}_j$ measurements on her subsystem labeled by $x_j$, with each measurement producing ${O}_j$ possible outcomes labeled by $a_j$ ($j = 1, 2, \dots, m$). Trusted parties are represented as $\mathcal{B}_k$ ($k = 1, 2, \dots, n$). All untrusted (trusted) parties are collected as $\bar{\mathcal{A}} = (\mathcal{A}_1, \mathcal{A}_2, \dots, \mathcal{A}_m)$ [$\bar{\mathcal{B}} = (\mathcal{B}_1, \mathcal{B}_2, \dots, \mathcal{B}_n)$]. We introduce the incidence matrix~\cite{rossetUniversalBoundCardinality2018}, defined as $I \in \{0, 1\}^{l \times m}$ ($\hat{I} \in \{0, 1\}^{l \times n}$), where $I_{i,j} = 1$ ($\hat{I}_{i,k} = 1$) if the $i$-th source is connected to the $j$-th untrusted party ($k$-th trusted party). To ensure that the scenarios are well-defined, we make the following reasonable assumption: for all $i$, if $\sum_k \hat{I}_{i,k} \neq 0$, then it must follow that $\sum_j I_{i,j} \neq 0$.
	
	In the network steering scenarios, after all untrusted parties $\bar{\mathcal{A}}$ complete their measurements, the resulting subnormalized states for the trusted parties $\bar{\mathcal{B}}$ are given by
	\begin{equation}\label{equ1}
		\sigma_{\bar{a}|\bar{x}}^{\bar{\mathcal{B}}} = 
		\mathrm{Tr}_{\bar{\mathcal{A}}} \Bigg[ 
		\bigg( \bigotimes_{j=1}^{m} M_{a_j|x_j}^{\mathcal{A}_j} \bigg) 
		\otimes 
		\bigg( \bigotimes_{k=1}^{n} \mathcal{I}^{\mathcal{B}_k} \bigg) 
		\cdot 
		\bigotimes_{i=1}^{l} \rho_i
		\Bigg],
	\end{equation}
	where $\bar{a} = (a_1, a_2, \dots, a_m)$, $\bar{x} = (x_1, x_2, \dots, x_m)$, $\mathrm{Tr}_{\bar{\mathcal{A}}}$ denotes taking the partial trace over all subsystems of $\bar{\mathcal{A}}$, and $M_{a_j|x_j}^{\mathcal{A}_j}$ are positive operators satisfying $\sum_{a_j} M_{a_j|x_j}^{\mathcal{A}_j} = \mathcal{I}^{\mathcal{A}_j}$ for all $x_j$. 
	
	The set of subnormalized states $\{\sigma_{\bar{a}|\bar{x}}^{\bar{\mathcal{B}}}\}_{\bar{a}, \bar{x}}$ is commonly referred to as a network assemblage. If this network assemblage can be decomposed as
	\begin{equation}\label{equ2}
		\sigma_{\bar{a}|\bar{x}}^{\bar{\mathcal{B}}} = \prod_{i=1}^{l} \sum_{\lambda_i} P_i(\lambda_i) \prod_{j=1}^{m} P_{\mathcal{A}_j}(a_j|x_j\lambda_{[j]}) \bigotimes_{k=1}^{n} \sigma_{\lambda_{[k]}}^{\mathcal{B}_k},
	\end{equation}
	then it can be concluded that the system admits a NLHS model. Here, $\lambda_i$ is a local hidden variable sampled from the value set $\Omega_i$ with probability  distribution $P_i(\lambda_i)$, $P_{\mathcal{A}_j}(a_j|x_j\lambda_{[j]})$ represents the local response function of $\mathcal{A}_j$, $\lambda_{[j]} = \{\lambda_i | I_{i,j} = 1\}$, and $\sigma_{\lambda_{[k]}}^{\mathcal{B}_k}$ denotes the local hidden state of $\mathcal{B}_k$. Conversely, if the network assemblage $\{\sigma_{\bar{a}|\bar{x}}^{\bar{\mathcal{B}}}\}_{\bar{a}, \bar{x}}$ does not admit the NLHS model, it demonstrates the existence of network steering. In other words, $\bar{\mathcal{A}}$ are capable of remotely steering the subsystems of $\bar{\mathcal{B}}$.
	
	In the following, we take the bilocal network scenario as an example and provide a detailed definition of network steering in this scenario. As shown in Fig.~\hyperref[fig1]{\ref{fig1}(\subref{fig1-(a)})}, where the central party $\mathcal{A}$ is untrusted and the endpoint parties 	 $\mathcal{B}_1,\mathcal{B}_2$ are trusted. $\mathcal{A}$ and $\mathcal{B}_1$ share the state $\rho_1$, while $\mathcal{A}$ and $\mathcal{B}_2$ share the state $\rho_2$.  $\mathcal{A}$ performs a fixed measurement $M_{a}^{\mathcal{A}}$, the subnormalized states resulting from this measurement are
	\begin{equation}\label{equ3}
		\sigma_{a}^{\bar{\mathcal{B}}} = \mathrm{Tr}_{{\mathcal{A}}} \left[ \left(\mathcal{I}^{\mathcal{B}_1} \otimes M_{a}^{\mathcal{A}}  \otimes \mathcal{I}^{\mathcal{B}_2} \right) \rho_1\otimes\rho_2 \right],
	\end{equation}
	where $\bar{\mathcal{B}}=(\mathcal{B}_1,\mathcal{B}_2)$. The set of subnormalized states $\{\sigma_{a}^{\bar{\mathcal{B}}}\}_{a}$ is referred to as the network assemblage. If this assemblage can be decomposed as
	\begin{equation}\label{equ4}
		\sigma_{a}^{\bar{\mathcal{B}}}=  \sum_{\lambda_1,\lambda_2} P_1(\lambda_1)P_2(\lambda_2) P_{\mathcal{A}}(a|\lambda_{1},\lambda_{2}) \sigma_{\lambda_1}^{\mathcal{B}_1}\otimes\sigma_{\lambda_2}^{\mathcal{B}_2},
	\end{equation}
	then we deduce that the NLHS model exists. Otherwise, this indicates the presence of network steering from the central party to the endpoint parties.
	
	In addition, the general definition introduced above also applies to standard steering scenarios [see Figs.~\hyperref[fig1]{\ref{fig1}(\subref{fig1-(b)})}-\hyperref[fig1]{\ref{fig1}(\subref{fig1-(d)})} for examples]. In the next section, we will provide a detailed explanation of the method used to measure quantum steerability in networks.

\section{\label{sec3:level1} METHOD FOR MEASURING NETWORK QUANTUM STEERABILITY}

	In the study of network quantum steering, we address the fundamental problem of identifying, for a given network assemblage, the closest assemblage that admits the NLHS model. The distance between these two assemblages acts as a quantitative measure of network quantum steerability. 
	
	To approach this problem systematically, we first introduce a metric analogous to that employed in Ref. \cite{kuEinsteinPodolskyRosenSteeringIts2018}, which has been rigorously established as a valid measure of quantum steerability. Given two distinct assemblages, $\{\sigma_{\bar{a}|\bar{x}}\}_{\bar{a}, \bar{x}}$ and $\{\sigma_{\bar{a}|\bar{x}}^*\}_{\bar{a}, \bar{x}}$, the distance between them can be defined as
	\begin{equation}\label{equ5}
		\normalsize
		\mathcal{D}_{\mathrm{A}}(\{\sigma_{\bar{a}|\bar{x}}\}_{\bar{a}, \bar{x}},\{\sigma_{\bar{a}|\bar{x}}^*\}_{\bar{a}, \bar{x}}) =\frac{1}{ {\textstyle \prod_{j=1}^{m}}  N_j 	} \sum_{\bar{a}, \bar{x}}  \mathcal{D}_{\mathrm{T}}(\sigma_{\bar{a}|\bar{x}},\sigma_{\bar{a}|\bar{x}}^*),
	\end{equation}
	where $\mathcal{D}_{\mathrm{T}}$ denotes the trace distance between quantum states, defined as $\mathcal{D}_{\mathrm{T}}(\sigma,\sigma^{*}) := \frac{1}{2}\|\sigma-\sigma^{*}\|_1$ with  $\|X\|_1 := \mathrm{Tr}(\sqrt{X^\dagger X})$ is the trace norm.
	
	\begin{figure*}[ht!]
		\centering
		\includegraphics[width=16.5cm]{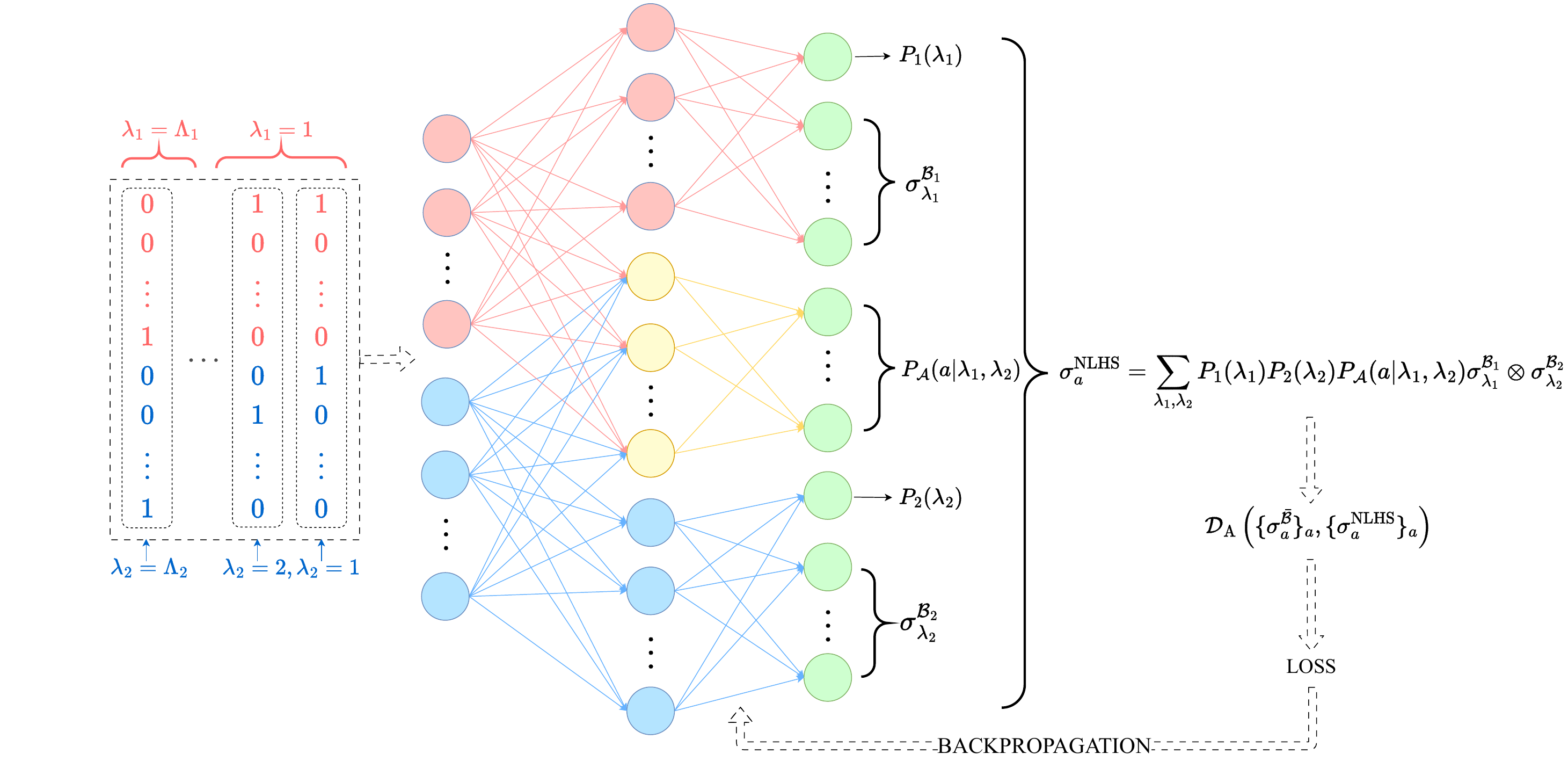}
		\caption{Diagram of modeling causal information with ANNs. The modeling process is generally organized into three distinct stages: the input stage, the causal propagation stage, and the optimization stage. In the input stage, data points from the input set $\Omega_\mathrm{IN}$ are fed into the ANN. The total number of data points in $\Omega_\mathrm{IN}$ determines the batch size, with a batch being completed when all data points are processed by the ANN. The causal propagation stage demonstrates the encoding and transmission of causal information through the ANN. In this stage, pink neurons are influenced by the hidden variable $\lambda_1$, blue neurons by $\lambda_2$, and yellow neurons by both $\lambda_1$ and $\lambda_2$. The output layer, represented by the green neurons, generates outputs that serve as optimization variables in Eq.~\eqref{equ6}. Finally, the optimization stage illustrates the procedure in which the ANN computes the loss after processing each batch, calculates gradients through backpropagation, and updates the model parameters accordingly.}
		\label{fig2}
	\end{figure*}
	
	Next, we will illustrate our method with the bilocal network steering scenario introduced in Sec.~\ref{sec2:level1} as a case study. Notably, our approach exhibits high generality and can be readily adapted to accommodate any quantum network with appropriate modifications.
	
	The principal objective of this work is to determine the network assemblage $\{\sigma_{a}^{\mathrm{NLHS}}\}_{a}$ that most closely approximates a given network assemblage $\{\sigma_{a}^{\bar{\mathcal{B}}}\}_{a}$, subject to the constraints imposed by the NLHS model. Accordingly, the problem can be cast as a nonlinear optimization problem, formulated as
	\begin{equation}\label{equ6}
		\begin{aligned} \mathcal{S}^{\mathrm{N}} := \min &\  \mathcal{D}_{\mathrm{A}} (\{\sigma_{a}^{\bar{\mathcal{B}}}\}_{a},\{\sigma_{a}^{\mathrm{NLHS}}\}_{a}),\\
			\mathrm{s.t.}&\  \{\sigma_{a}^{\mathrm{NLHS}}\}_{a} \in \mathrm{NLHS} ,\\&\ \  \sum_a \sigma^{\mathrm{NLHS}}_{a} =\rho^{\mathcal{\bar{\mathcal{B}}}},\end{aligned}
	\end{equation}
	where the first constraint signifies that the network assemblage $\{\sigma^{\text{NLHS}}_{a}\}_{a}$ admits the $\mathrm{NLHS}$ model, and the second constraint represents the consistency condition introduced by Cavalcanti \textit{et al.} \cite{cavalcantiQuantitativeRelationsMeasurement2016}, which ensures that the network assemblage $\{\sigma_a^\mathrm{NLHS}\}_a$ defines the same reduced state as $\{\sigma_a^{\bar{\mathcal{B}}}\}_a$, i.e., $\sum_a \sigma_a^\mathrm{NLHS} = \sum_a \sigma_a^{\bar{\mathcal{B}}} = \rho^{\bar{\mathcal{B}}}$. $\mathcal{S}^{\mathrm{N}}$ is referred to as the measure of network quantum steerability.
	
	Tackling Eq.~\eqref{equ6} entails considerable difficulty owing to the intricate presence of multiple nonlinear terms, rendering traditional optimization techniques ineffective, especially as the quantum network grows in complexity.  Ref.~\cite{kuEinsteinPodolskyRosenSteeringIts2018} investigated a related problem in the bipartite steering scenario, but their approach was limited to computing the lower and upper bounds of the original problem. In contrast, inspired by  
	Ref.~\cite{krivachyNeuralNetworkOracle2020}, we directly model causal information using ANNs to solve Eq.~\eqref{equ6}. This approach is especially advantageous, as ANNs are inherently structured as directed acyclic graphs, which closely mimic the topology of quantum networks. 
	
	In network scenarios, we assume that each source operates independently, ensuring no interference between their behaviors. Within the local model, each source aims to control the output corresponding to each input of the untrusted party it is connected to via classical information $\lambda_i$ \cite{puseyNegativitySteeringStronger2013, brunnerBellNonlocality2014}. This implies that in finite-dimensional spaces, if the number of inputs and outputs for the untrusted party is finite, the cardinality of $\Omega_i$ (denoted as $\Lambda_i$) must also be finite. Without loss of generality, we assume that ${\Omega_i} = \{1, 2, \dots, \Lambda_i\}$. 
	
	To solve Eq.~\eqref{equ6} using an ANN-based method, we construct the input set in accordance with the principle of causality \cite{woodLessonCausalDiscovery2015,fritzBellsTheoremII2016} as $\Omega_{\mathrm{IN}} = \Omega_1 \times \Omega_2$ (with $\times$ denoting the Cartesian product). Each value of a hidden variable is encoded using a one-hot representation (as illustrated on the left side of Fig.~\ref{fig2}). Furthermore, all variables in Eq.~\eqref{equ6} are affected by the hidden variables $\lambda_1$ and/or $\lambda_2$. Consequently, all optimization variables, including $P_{1}(\lambda_1)$, $P_2(\lambda_2)$, $P_{\mathcal{A}}(a|\lambda_1, \lambda_2)$, $\sigma_{\lambda_1}^{\mathcal{B}_1}$, and $\sigma_{\lambda_2}^{\mathcal{B}_2}$, are represented as outputs of the ANN. The constraints and objective function specified in Eq.~(\ref{equ6}) are integrated into the ANN as its loss function. Since certain outputs of the ANN correspond to probability values, the activation function of the final layer is set to sigmoid. For local hidden states $\sigma_\lambda$, which can be regarded as pure states in finite dimensions, some outputs of the ANN are interpreted as state vectors. Specifically, a state vector is represented as $|\psi\rangle = \sum_{t=0}^{d-1} \alpha_t |t\rangle$, where $\{|t\rangle\}_{t=0}^{d-1}$ constitutes the computational basis in $\mathbb{C}^d$, and the coefficients $\alpha_t(\in \mathbb{C})$ satisfy the normalization condition $\sum_{t=0}^{d-1} |\alpha_t|^2 = 1$. These coefficients are expressed as $\alpha_t = b_t + c_t\mathrm{i}$, where $b_t, c_t (\in \mathbb{R})$ correspond to the ANN's outputs. 
	
	The entire modeling process of the ANN is visualized in Fig.~\ref{fig2}. Moreover, since measurement choices also serve as “causes” within the quantum network, for scenarios involving more than one measurement setting, they can similarly be incorporated as inputs to the ANN. It is worth emphasizing that, on the one hand, our approach subtly harnesses the causal information intrinsic to the quantum network, thereby enabling seamless integration with ANNs. On the other hand, our approach adopts a generative method wherein the network assemblage $\{\sigma_a^\mathrm{NLHS}\}_a$ that satisfies the NLHS model can be dynamically adjusted in response to the outcomes derived from the loss function. 

\section{\label{sec4:level1}NUMERICAL VALIDATION AND CASE EXPLORATION}

	To assess the validity of the proposed method, we first apply it to both bipartite and multipartite steering scenarios, comparing the results with several well-established conclusions. These comparisons confirm the accuracy and efficiency of our approach. Subsequently, we focus on the bilocal network steering scenario as a case exploration, deriving several novel results.
	
	\subsection{\label{sec4:level2-1} Measure of bipartite quantum steerability}
	
	\begin{figure*}[ht!]
		\includegraphics[width=\textwidth]{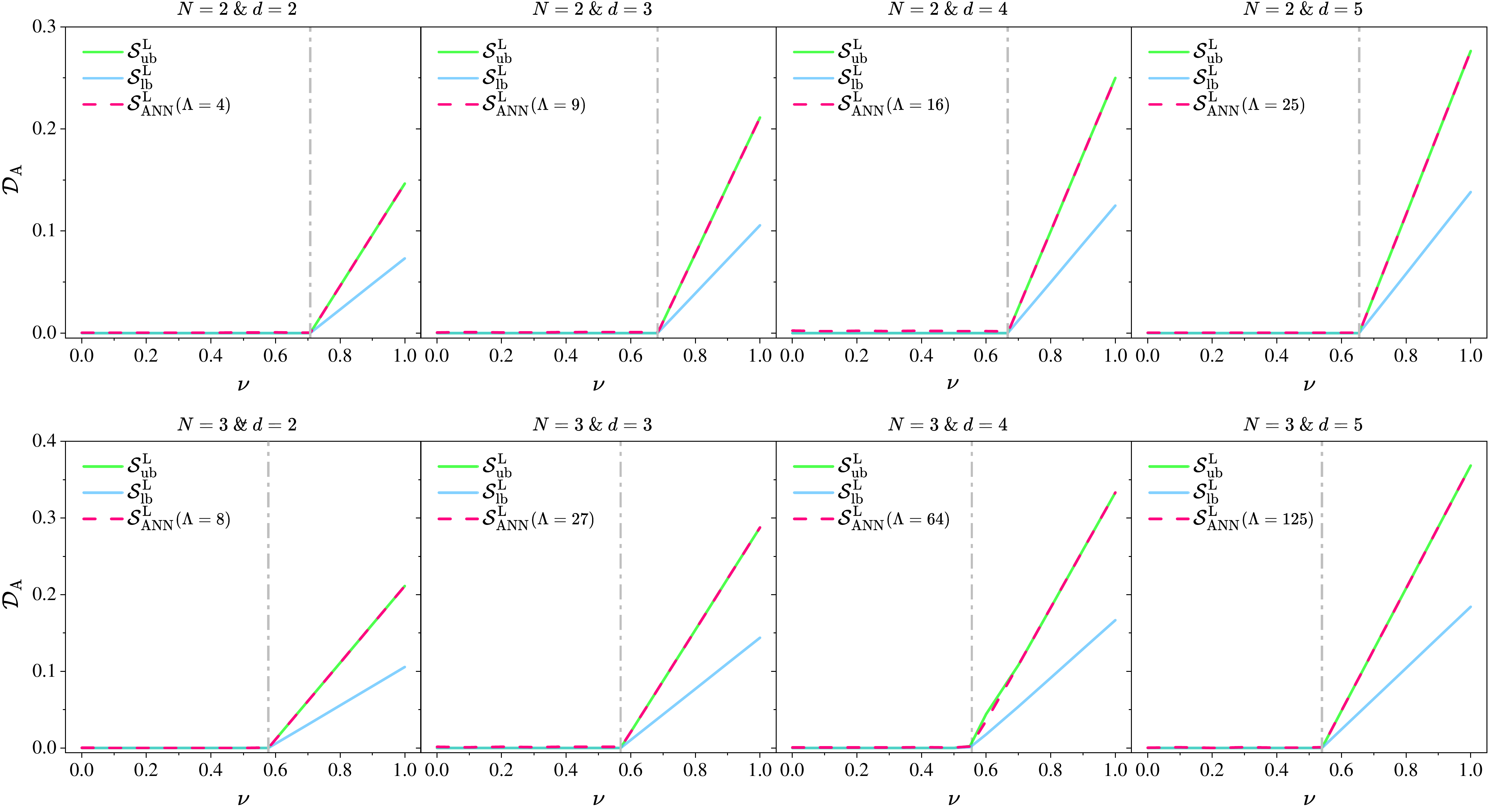}
		\caption{Numerical validation results in the bipartite steering scenario. We employ our ANN-based method to compute $\mathcal{S}^{\mathrm{L}}$ [Eq.~\eqref{equ8}] across diverse measurement settings and dimensions, obtaining the corresponding values $\mathcal{S}_{\mathrm{ANN}}^{\mathrm{L}}$. To rigorously evaluate the validity of our approach, we derive the theoretical bounds $\mathcal{S}_{\mathrm{lb}}^{\mathrm{L}}$ and $\mathcal{S}_{\mathrm{ub}}^{\mathrm{L}}$ for $\mathcal{S}^{\mathrm{L}}$ using the method introduced in Ref. \cite{kuEinsteinPodolskyRosenSteeringIts2018}, which provides a reliable benchmark for comparison. The dash-dot lines indicate the steering thresholds, with the exact values presented in Table~\ref{tab1}.}
		\label{fig3}
	\end{figure*}
	
	In the bipartite steering scenario, we focus on isotropic states, a class of quantum states whose steering properties have been extensively studied~\cite{uolaQuantumSteering2020,cavalcantiExperimentalCriteriaSteering2009,cavalcantiQuantitativeRelationsMeasurement2016}. These states are formally defined as
	\begin{equation}\label{equ7}
		\rho_{\mathrm{Iso}} (d,\nu) =\nu|\Phi_d^+\rangle\langle \Phi_d^+| +(1-\nu)\frac{\mathcal{I}}{d^2},
	\end{equation}
	where $\nu$ is the visibility satisfying $ \nu \in [0,1]$, $d$ represents the dimension of the Hilbert space, $ |\Phi_d^+\rangle = (1/\sqrt{d})\sum_{t=0}^{d-1}|t,t\rangle $ is the maximally entangled state, and ${\mathcal{I}}/{d^2}$ corresponds to the two-qudits maximally mixed state. 
	
	The untrusted party $\mathcal{A}$ performs measurements based on mutually unbiased bases (MUBs) \cite{schwingerUnitaryOperatorBases1960}, hereafter referred to as mutually unbiased measurements. Formally, two orthonormal bases, $\{ |\beta_s\rangle \}_{s=0}^{d-1}$ and $\{ |\gamma_t\rangle \}_{t=0}^{d-1}$, are called mutually unbiased if and only if $|\langle \beta_s | \gamma_t \rangle|^2 = {1}/{d}$ for all $s$ and $t$. In this work, we follow the construction method for MUBs described in Ref. \cite{klappenecker2004mub}. 
	
	Analogous to Eq.~\eqref{equ6}, the optimization model adopted in this scenario is defined by
	\begin{equation}\label{equ8}
		\begin{aligned}\mathcal{S}^{\mathrm{L}}:= \min &\    \mathcal{D}_{\mathrm{A}}(\{\sigma_{a|x}^{\mathcal B}\}_{a,x},\{ \sigma^{\text{LHS}}_{a|x}\}_{a,x}),\\
		\mathrm{s.t.}&\  \{ \sigma^{\text{LHS}}_{a|x}\}_{a,x} \in \mathrm{LHS},\\&\  \sum_a \sigma^{\text{LHS}}_{a|x} =\rho^{\mathcal B} \ \  \forall x, \end{aligned}
	\end{equation}
	where the first constraint signifies that the assemblage $\{\sigma^{\text{LHS}}_{a|x}\}_{a,x}$ admits the $\mathrm{LHS}$ model, the second constraint similarly represents the consistency condition, and $\mathcal{S}^{\mathrm{L}}$ is termed the bipartite quantum steerability. 
	
	To perform numerical validation, we adapt the ANN architecture described in Sec.~\ref{sec3:level1} to model the causal information pertinent to this scenario. Our analysis focuses on dimensions $d \leqslant 5$, where $\mathcal{A}$ performs ${N} = 2$ $({N} = 3)$ mutually unbiased measurements constructed from the first two (three) MUBs. For the special case of $d = 2$, the measurements performed by $\mathcal{A}$ correspond to the Pauli operators $Z$ and $X$ for ${N} = 2$, and $Z$, $X$, and $Y$ for ${N} = 3$. As in Ref.~\cite{puseyNegativitySteeringStronger2013}, the cardinality of $\Omega$ is set to $\Lambda = d^{N}$. Additionally, to facilitate comparative analysis, we derive the lower and upper bounds of $\mathcal{S}^{\mathrm{L}}$ (denoted as $\mathcal{S}_{\mathrm{lb}}^{\mathrm{L}}$ and $\mathcal{S}_{\mathrm{ub}}^{\mathrm{L}}$) using the approach introduced in Ref.~\cite{kuEinsteinPodolskyRosenSteeringIts2018}.
	
	\begin{table*}[ht!]
		\caption{\label{tab1} The steering threshold and the analytical expression of bipartite quantum steerability for each case, with $\nu \in [0,1]$. }
		\renewcommand{\arraystretch}{1.7} 
		\setlength{\tabcolsep}{5pt} 
		\begin{ruledtabular}
			\begin{tabular}{cc|cccc}
				& $d$ & 2 & 3 & 4 & 5 \\ \hline 
				\multirow{2}{*}{${N} = 2$}&$\nu_s$  
				& $0.7071 \approx \frac{1}{\sqrt{2}}$ 
				& $0.6830 \approx \frac{1 + \sqrt{3}}{4}$
				& $0.6667 \approx \frac{2}{3}$ 
				& $0.6545 \approx \frac{3+\sqrt{5}}{8}$ \\ 
				&$\mathcal{S}^{\mathrm{L}}$  
				& $\max\left\{0,\frac{1}{2}\left(\nu - \frac{1}{\sqrt{2}}\right)\right\}$ 
				& $\max\left\{0,\frac{2}{3}\left(\nu - \frac{1+\sqrt{3}}{4}\right)\right\}$ 
				& $\max\left\{0,\frac{3}{4}\left(\nu - \frac{2}{3}\right)\right\}$
				& $\max\left\{0,\frac{4}{5}\left(\nu - \frac{3+\sqrt{5}}{8}\right)\right\}$ \\ 
				\multirow{2}{*}{${N} = 3$}&$\nu_s$ 
				& $0.5774 \approx \frac{1}{\sqrt{3}}$ 
				& $0.5686 \approx \frac{\cos(\pi/8)}{\sqrt{3}}$ 
				& $0.5556 \approx \frac{5}{9}$ 
				& $0.5393 \approx \frac{1+\sqrt{5}}{6}$ \\ 
				&$\mathcal{S}^{\mathrm{L}}$ 
				& $\max\left\{0,\frac{1}{2}\left(\nu - \frac{1}{\sqrt{3}}\right)\right\}$ 
				& $\max\left\{0,\frac{2}{3}\left(\nu - \frac{\cos(\pi/8)}{\sqrt{3}}\right)\right\}$ 
				& $\max\left\{0,\frac{3}{4}\left(\nu - \frac{5}{9}\right)\right\}$ 
				& $\max\left\{0,\frac{4}{5}\left(\nu - \frac{1+\sqrt{5}}{6}\right)\right\}$ \\ 
			\end{tabular}
		\end{ruledtabular}
	\end{table*}
	
	Figure \ref{fig3} presents our numerical results. For dimension $d=2$, the steering thresholds $\nu_s$—defined as the visibility values above which the existence of steering can be demonstrated—are consistent with the results reported in Refs.~\cite{cavalcantiExperimentalCriteriaSteering2009,cavalcantiQuantitativeRelationsMeasurement2016}. Furthermore, across all examined cases, $\mathcal{S}_{\mathrm{ANN}}^{\mathrm{L}}$ remains within the theoretical bounds $[\mathcal{S}_{\mathrm{lb}}^{\mathrm{L}}, \mathcal{S}_{\mathrm{ub}}^{\mathrm{L}}]$, with $\mathcal{S}_{\mathrm{ANN}}^{\mathrm{L}}$ approximating twice the value of $\mathcal{S}_{\mathrm{lb}}^{\mathrm{L}}$ and being close to $ \mathcal{S}_{\mathrm{ub}}^{\mathrm{L}} $. To further demonstrate the strengths of the method, we analyze the special case where ${N} = d = 2$. Leveraging the numerical results obtained from the ANN, we explicitly reconstruct the mathematical expression of the corresponding LHS model,
	\begin{subequations}\label{equ9}
		\begin{equation}\label{equ9-a}
			\small
			P(\lambda) = \left \{\frac{1}{4}, \frac{1}{4}, \frac{1}{4}, \frac{1}{4}\right\},
		\end{equation}
		\begin{align}\label{equ9-b}
			\small
			P_{\lambda}(a,x) =  
			&\left\{
			\begin{bmatrix}
				1&1 \\
				0&0
			\end{bmatrix},\begin{bmatrix}
				1 & 0\\
				0 & 1
			\end{bmatrix},\begin{bmatrix}
				0 & 1\\
				1 & 0
			\end{bmatrix},\begin{bmatrix}
				0 & 0\\
				1 & 1
			\end{bmatrix}
			\right\},
		\end{align}
		
		\begin{align}\label{equ9-c}
			\small
			\sigma_{\lambda}^{\mathcal{B}} =  
			&\left\{
			\begin{bmatrix}
				\frac{1}{2}+ \frac{\nu}{2} & \frac{\nu}{2} \\
				\frac{\nu}{2} &  \frac{1}{2}- \frac{\nu}{2}
			\end{bmatrix}, 
			\begin{bmatrix}
				\frac{1}{2}- \frac{\nu}{2} & \frac{\nu}{2} \\
				\frac{\nu}{2} &  \frac{1}{2}+ \frac{\nu}{2}
			\end{bmatrix}, \right. \nonumber \\ 
			&\ \ \left.
			\begin{bmatrix}
				\frac{1}{2}+ \frac{\nu}{2} & -\frac{\nu}{2} \\
				-\frac{\nu}{2} &  \frac{1}{2}- \frac{\nu}{2}
			\end{bmatrix}, 
			\begin{bmatrix}
				\frac{1}{2}- \frac{\nu}{2} & -\frac{\nu}{2} \\
				-\frac{\nu}{2} &  \frac{1}{2}+ \frac{\nu}{2}
			\end{bmatrix}
			\right\}.
		\end{align}
	\end{subequations}
	Notably, for any $\lambda$, $\sigma_{\lambda}^{\mathcal{B}}$ is positive semidefinite only when $\nu \in  [0, {1}/{\sqrt{2}}]$, which allows for the explicit construction of the LHS model. For $\nu \in ({1}/{\sqrt{2}}, 1]$, it is observed from the numerical results that the assemblage $\{ \sigma^{\text{LHS}}_{a|x}\}_{a,x}$ closest to the given assemblage remains described by Eq.~\eqref{equ9}. However, a key difference is that the visibility $\nu$ in Eq.~\eqref{equ9-c} is fixed at ${1}/{\sqrt{2}}$. Building upon this, we can derive the analytical expression for $\mathcal{S}^{\mathrm{L}}$, i.e.,
	\begin{equation}\label{equ10}
		\mathcal{S}^{\mathrm{L}} = \max\left\{0,\frac{1}{2}\left(\nu - \frac{1}{\sqrt{2}}\right)\right\},
	\end{equation}
	where $\nu \in [0,1]$. The above results, which align with the geometric analysis in Ref.~\cite{kuEinsteinPodolskyRosenSteeringIts2018}, further substantiate our approach from a different perspective. In Table \ref{tab1}, we tabulate the analytical expressions for $\mathcal{S}^{\mathrm{L}}$ across all cases, derived using the same method. We can observe that, for a fixed system dimension, increasing the number of measurement settings allows for the extraction of more information, thereby enabling the detection of steering at smaller values of \(\nu\). Additionally, for  $\mathcal{S}^\mathrm{L} > 0$, the slope $p$ appears to follow the relation $p = (d-1)/d$.
	
	\subsection{\label{sec4:level2-2} Measure of multipartite quantum steerability}
	
	We next validate our method within the multipartite steering scenarios, focusing primarily on the following tripartite states
	\begin{equation}\label{equ11}
		\rho_{\mathrm{GHZ}}(\nu) = \nu|\mathrm{GHZ}\rangle\langle\mathrm{GHZ}|+(1-\nu)\frac{\mathcal{I}}{8},
	\end{equation}
	where $|\mathrm{GHZ}\rangle = (|000\rangle + |111\rangle)/\sqrt{2}$. The multipartite steering properties of the states in Eq.~\eqref{equ11} have been investigated in Ref.~\cite{cavalcantiDetectionEntanglementAsymmetric2015}, primarily focusing on the detection of the multipartite steering thresholds through semidefinite programming (SDP). In this study, however, we explore the steerability of these states by employing ANNs. For clarity, we denote the local model for a single untrusted party (1-UNT) as ${\mathrm{MLHS_1}}$, which satisfies Eq.~\eqref{equ2} with $l=m=1$ and $n=2$. For the scenario involving two untrusted parties (2-UNT), the local model is denoted as ${\mathrm{MLHS_2}}$, which satisfies Eq.~\eqref{equ2} with $l=n=1$ and $m=2$.
	
	In the 1-UNT scenario [see Fig.~\hyperref[fig1]{\ref{fig1}(\subref{fig1-(c)})}], where only $\mathcal{A}$ is untrusted, the party performs either two mutually unbiased measurements ($Z$ and $X$) or three mutually unbiased measurements ($Z$, $X$, and $Y$), corresponding to $N=2$ or $N=3$, respectively. The associated optimization model is given by
	\begin{equation}\label{equ12}
		\begin{aligned}\mathcal{S}^{\mathrm{M}_1}:= \min&\ \mathcal{D}_{\mathrm{A}}(\{\sigma_{a|x}^{\bar{\mathcal{B}}}\}_{a,x},\{ \sigma^{\mathrm{MLHS}_1}_{a|x}\}_{a,x}),\\
			\mathrm{s.t.}&\  \{\sigma^{\mathrm{MLHS_1}}_{a|x}\}_{a,x} \in \mathrm{MLHS}_1,\\&\  \sum_a \sigma^{\mathrm{MLHS_1}}_{a|x} =\rho^{\bar{\mathcal{B}}} \ \  \forall x, \end{aligned}
	\end{equation}
	where $\mathcal{S}^{\mathrm{M}_1}$ is termed the measure of multipartite quantum steerability in the presence of a single untrusted party.
	
	For the 2-UNT scenario [see Fig.~\hyperref[fig1]{\ref{fig1}(\subref{fig1-(d)})}], where both $\mathcal{A}_1$ and $\mathcal{A}_2$ are untrusted, it is assumed that they have the same measurement settings, with each performing either two mutually unbiased  measurements ($Z$ and $X$) or three mutually unbiased measurements ($Z$, $X$, and $Y$), corresponding to $N_1 = N_2 = 2$ or $N_1 = N_2 = 3$, respectively. This model assumes the form
	\begin{equation}\label{equ13}
		\begin{aligned}\mathcal{S}^{\mathrm{M}_2}:= \min&\ \mathcal{D}_{\mathrm{A}}(\{\sigma_{\bar{a}|\bar{x}}^{\mathcal B}\}_{\bar{a},\bar{x}},\{ \sigma^{\mathrm{MLHS}_2}_{\bar{a}|\bar{x}}\}_{\bar{a},\bar{x}}),\\
			\mathrm{s.t.}&\  \{\sigma^{\text{MLHS}_2}_{\bar{a}|\bar{x}}\}_{\bar{a},\bar{x}} \in \mathrm{MLHS}_2 ,\\&\  \sum_{\bar{a}} \sigma^{\mathrm{MLHS}_2}_{\bar{a}|\bar{x}} =\rho^{\mathcal B} \ \  \forall \bar{x}. \end{aligned}
	\end{equation}
	Here, $\mathcal{S}^{\mathrm{M}_2}$ serves as the measure of multipartite quantum steerability with two untrusted parties.
	
	\begin{figure*}[ht!]
		\includegraphics[width=\textwidth]{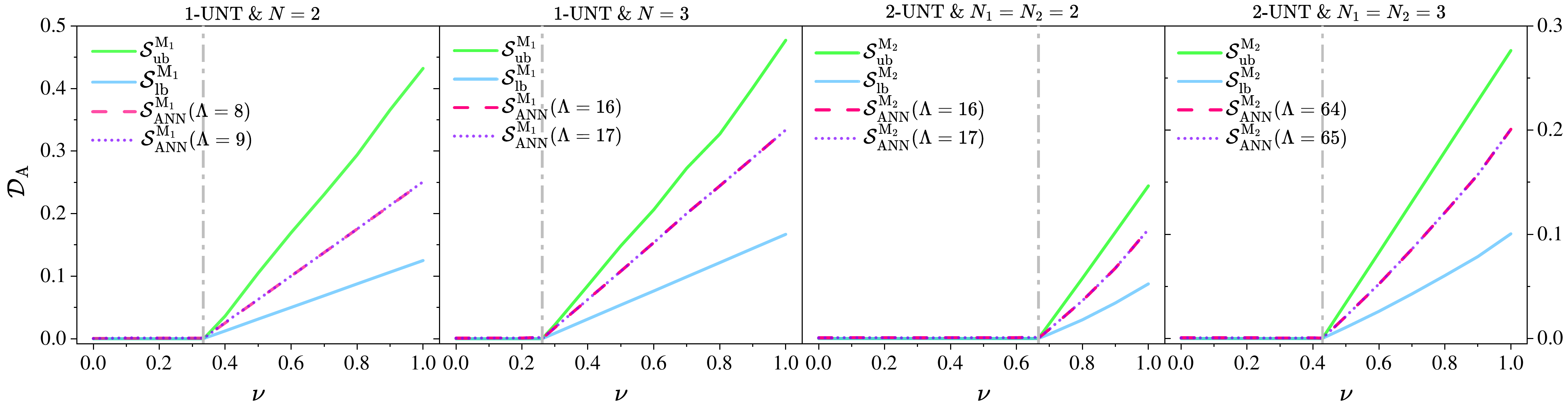}
		\caption{Numerical validation results in the multipartite steering scenarios. The multipartite steering thresholds, denoted by $\nu_{ms}$, take the values $0.3333 \approx {1}/{3}$, $0.2612 \approx {1}/{(2\sqrt{2}+1)}$, $0.6667 \approx {2}/{3}$, and $0.4286 \approx {3}/{7}$, respectively. }
		\label{fig4}
	\end{figure*}

	We validate our method in these scenarios and, for comparative analysis, derive the theoretical bounds for $\mathcal{S}^{\mathrm{M}_1}$ and $\mathcal{S}^{\mathrm{M}_2}$ using SDP techniques (cf. Appendix~\ref{app1:sec1}). The corresponding numerical results are plotted in Fig.~\ref{fig4}. In the 1-UNT scenario, the plots of $ \mathcal{S}_{\mathrm{ANN}}^{\mathrm{M}_1} $ exhibit numerical convergence when the cardinality of the hidden variable set satisfies $\Lambda \geqslant 2^{(N+1)}$. Similarly, in the 2-UNT scenario, convergence of $ \mathcal{S}_{\mathrm{ANN}}^{\mathrm{M}_2} $ is achieved when $\Lambda \geqslant 2^{(N_1+N_2)}$. For both $t = 1, 2$, $ \mathcal{S}_{\mathrm{ANN}}^{\mathrm{M}_{t}} $ consistently remains within the theoretical bounds $[\mathcal{S}_{\mathrm{lb}}^{\mathrm{M}_{t}}, \mathcal{S}_{\mathrm{ub}}^{\mathrm{M}_{t}}]$, further demonstrating the validity of our ANN-based method. Finally, it is worth noting that $ \mathcal{S}_{\mathrm{ANN}}^{\mathrm{M}_{t}} $ approximates twice the value of $ \mathcal{S}_{\mathrm{lb}}^{\mathrm{M}_{t}} $, while no longer being close to $ \mathcal{S}_{\mathrm{ub}}^{\mathrm{M}_{t}} $, reflecting the distinct differences between bipartite and multipartite quantum steerability.
	
	\subsection{\label{sec4:level2-3} Measure of network quantum steerability}
	
	In network steering scenarios, the presence of multiple independent sources—compared to the previously studied standard steering scenarios—introduces significant complexity in measuring network quantum steerability. In this subsection, we focus on the bilocal  network steering scenario as a case exploration and derive several novel results.
	
	As depicted in Fig.~\hyperref[fig1]{\ref{fig1}(\subref{fig1-(a)})}, we assume that $\mathcal{A}$ and $\mathcal{B}_1$, as well as $\mathcal{A}$ and $\mathcal{B}_2$, respectively share the states
	\begin{equation}\label{equ14}
		\begin{split}
			\rho_1 &= \nu|\Phi_2^+\rangle \langle \Phi_2^+| + (1-\nu)\frac{\mathcal{I}}{4}, \\
			\rho_2 &= \omega|\Phi_2^+\rangle \langle \Phi_2^+| + (1-\omega)\frac{\mathcal{I}}{4},
		\end{split}
	\end{equation}
	where $\nu$ and $\omega$ are the visibilities of the two states, distinct from each other and satisfying $\nu, \omega \in [0,1]$. $\mathcal{A}$ performs a BSM defined by the following operators
	\begin{equation}\label{equ15}
		\begin{split}
			M_{0}^{\mathcal{A}} = |\Phi_2^+\big>\big<\Phi_2^+\big|,\ M_{1}^{\mathcal{A}} = |\Phi_2^-\big>\big<\Phi_2^-\big|, \\
			M_{2}^{\mathcal{A}} = |\Psi_2^+\big>\big<\Psi_2^+\big|,\ 
			M_{3}^{\mathcal{A}} = |\Psi_2^-\big>\big<\Psi_2^-\big|,
		\end{split}
	\end{equation}
	where $|\Phi_2^+\rangle$, $|\Phi_2^-\rangle$, $|\Psi_2^+\rangle$, and $|\Psi_2^-\rangle$ are the four maximally entangled two-qubit Bell states .
	
	After $\mathcal{A}$ performs the BSM, the resulting network assemblage prepared for $\mathcal{B}_1$ and $\mathcal{B}_2$ is represented as $\{\sigma_{a}^{\bar{\mathcal{B}}}\}_{a=0}^3$, with explicit expressions given by
	\begin{equation}\label{equ16}
		\begin{split}
			\sigma_0^{\bar{\mathcal{B}}} = \frac{\nu\omega}{4}  |\Phi_2^+\big>\big<\Phi_2^+\big|+ \frac{1-\nu\omega}{16}\mathcal{I} , \\
			\sigma_1^{\bar{\mathcal{B}}} = \frac{\nu\omega}{4}  |\Phi_2^-\big>\big<\Phi_2^-\big|+ \frac{1-\nu\omega}{16}\mathcal{I} , \\
			\sigma_2^{\bar{\mathcal{B}}} = \frac{\nu\omega}{4}  |\Psi_2^+\big>\big<\Psi_2^+\big|+ \frac{1-\nu\omega}{16}\mathcal{I} , \\
			\sigma_3^{\bar{\mathcal{B}}} = \frac{\nu\omega}{4}  |\Psi_2^-\big>\big<\Psi_2^-\big|+ \frac{1-\nu\omega}{16}\mathcal{I} .
		\end{split}
	\end{equation}
	In this scenario, the entanglement of a single subnormalized state $\sigma_{a}^{\bar{\mathcal{B}}}$ signifies that the network assemblage $\{\sigma_{a}^{\bar{\mathcal{B}}}\}_{a=0}^3$ cannot satisfy the decomposition described by Eq.~\eqref{equ4}, indicating the presence of network steering from the central party to the endpoint parties \cite{jonesNetworkQuantumSteering2021}. This demonstrates that the existence of the NLHS model implies $\nu\omega \leqslant {1}/{3}$. 
	
\begin{figure}[t]
		\includegraphics[width=0.48\textwidth]{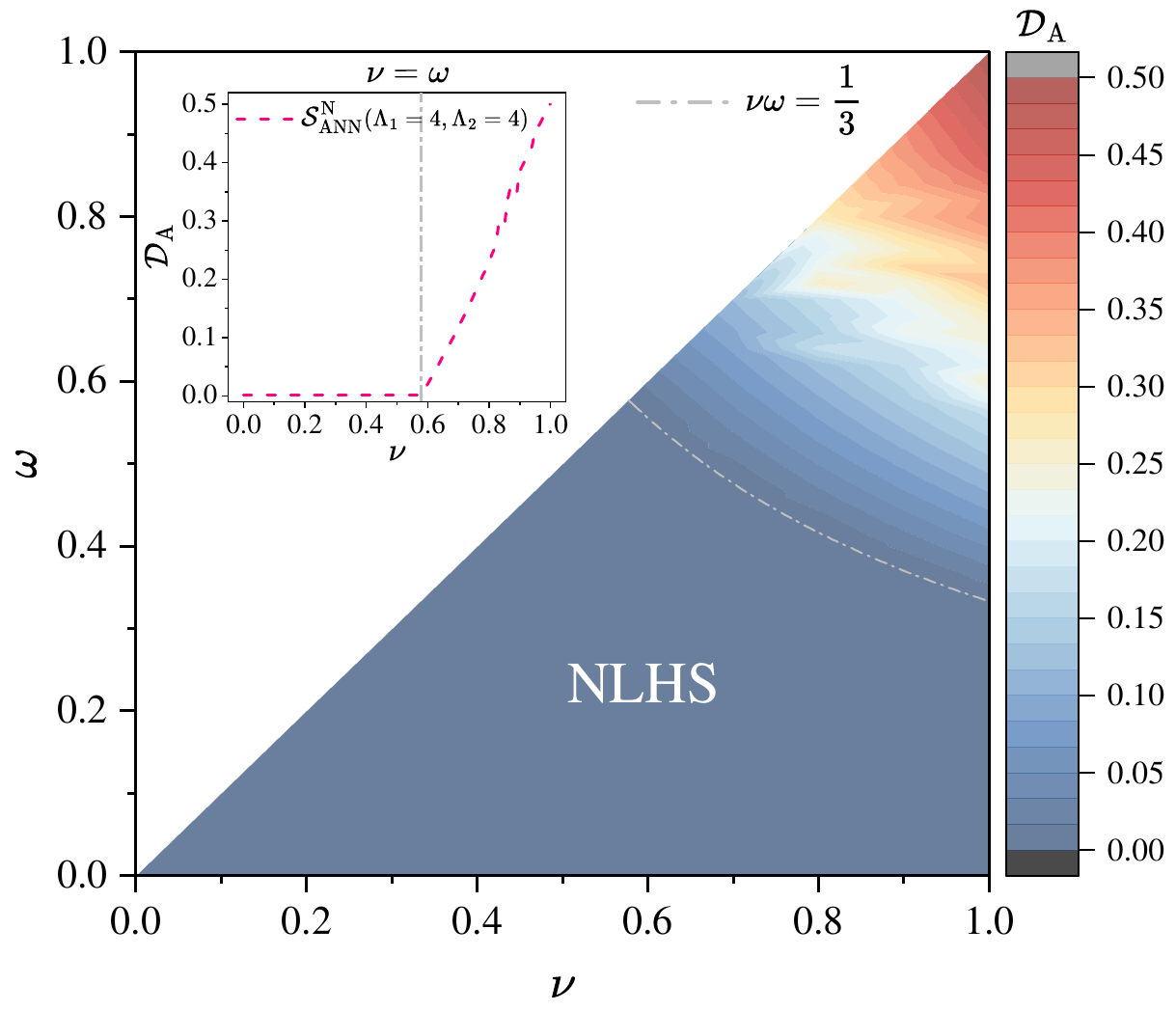}
		\caption{Network quantum steerability in the bilocal network steering scenario. The contour plot illustrates $\mathcal{S}_\mathrm{ANN}^\mathrm{N}$ as a function of the visibilities of the two shared quantum states in the network, $\nu$ (horizontal axis) and $\omega$ (vertical axis). The colormap represents the corresponding values of $\mathcal{S}_\mathrm{ANN}^\mathrm{N}$. Due to symmetry, only half of the plot is shown, specifically the region where $0 \leqslant \nu \leqslant 1$ and $0 \leqslant \omega \leqslant \nu$. Some oscillations (i.e., numerical artifacts) appear in the plot, which may result from insufficient iterations. The inset depicts $\mathcal{S}_\mathrm{ANN}^\mathrm{N}$ as a function of $\nu$ under the assumption $\nu = \omega$.}
		\label{fig5}
	\end{figure}

	According to Sec.~\ref{sec3:level1}, the optimization model employed in this scenario is given by Eq.~\eqref{equ6}, with the architecture of the ANN detailed in Fig.~\ref{fig2}, where we set $\Lambda_1 = \Lambda_2 = 4$. Numerical results from the ANN (see Fig.~\ref{fig5}) indicate that the curve $\nu \omega = {1}/{3}$ serves as the thresholds for network steering, i.e., network steering is absent when $\nu \omega \leqslant {1}/{3}$ and present otherwise.
	
	
	To further verify this conclusion, we reconstruct the NLHS model based on the numerical results, which take the following form
	\begin{subequations}\label{equ17}
		\label{}
		\begin{equation}\label{equ17-1}
			\small
			P_1(\lambda_1) = P_2(\lambda_2) = \left\{ \frac{1}{4},\frac{1}{4},\frac{1}{4},\frac{1}{4}\right\} , 
		\end{equation}
		\begin{equation}\label{equ17-2}
			\small
			\setlength{\arraycolsep}{2pt} 
			P_{\mathcal{A}}(a|\lambda_1,\lambda_2) = \! \left[ 
			\begin{array}{cccccccccccccccc} 
				0 & 0 & 0 & 1 & 1 & 0 & 0 & 0 & 0 & 0 & 1 & 0 & 0 & 1 & 0 & 0 \\ 
				0 & 0 & 1 & 0 & 0 & 1 & 0 & 0 & 0 & 0 & 0 & 1 & 1 & 0 & 0 & 0 \\ 
				1 & 0 & 0 & 0 & 0 & 0 & 0 & 1 & 0 & 1 & 0 & 0 & 0 & 0 & 1 & 0 \\ 
				0 & 1 & 0 & 0 & 0 & 0 & 1 & 0 & 1 & 0 & 0 & 0 & 0 & 0 & 0 & 1 
			\end{array} 
			\right]\!,
		\end{equation}
		
		\begin{align}\label{equ17-3}
			\small
			\sigma_{\lambda_1}^{\mathcal{B}_1}\! =\!  
			&\left\{\!
			\setlength{\arraycolsep}{2pt}
			\renewcommand{\arraystretch}{1.3} 
			\begin{bmatrix}
				\frac{1}{2} - \frac{\xi}{2} & \frac{\xi}{2} - \frac{\xi\mathrm{i}}{2} \\
				\frac{\xi}{2} + \frac{\xi \mathrm{i}}{2} & \frac{1}{2} + \frac{\xi}{2}
			\end{bmatrix}\!,\!
			\begin{bmatrix}
				\frac{1}{2} + \frac{\xi}{2} & \frac{\xi}{2} + \frac{\xi \mathrm{i}}{2} \\
				\frac{\xi}{2} - \frac{\xi \mathrm{i}}{2} & \frac{1}{2} - \frac{\xi}{2}
			\end{bmatrix}\!,\! \right. \nonumber \\ 
			&\ \ \left.
			\renewcommand{\arraystretch}{1.3}\!
			\setlength{\arraycolsep}{2pt}
			\begin{bmatrix}
				\frac{1}{2} - \frac{\xi}{2} & -\frac{\xi}{2} + \frac{\xi \mathrm{i}}{2} \\
				-\frac{\xi}{2} - \frac{\xi \mathrm{i}}{2} & \frac{1}{2} + \frac{\xi}{2}
			\end{bmatrix}\!,\!
			\begin{bmatrix}
				\frac{1}{2} + \frac{\xi}{2} & -\frac{\xi}{2} - \frac{\xi \mathrm{i}}{2} \\
				-\frac{\xi}{2} + \frac{\xi \mathrm{i}}{2} & \frac{1}{2} - \frac{\xi}{2}
			\end{bmatrix}\!
			\right\}\!,
		\end{align}
		\begin{align}\label{equ17-4}
			\small
			\sigma_{\lambda_2}^{\mathcal{B}_2} \!=\!  
			&\left\{\!
			\setlength{\arraycolsep}{2pt}
			\renewcommand{\arraystretch}{1.3} 
			\begin{bmatrix}
				\frac{1}{2} + \frac{\zeta}{2} & \frac{\zeta}{2} - \frac{\zeta \mathrm{i}}{2} \\
				\frac{\zeta}{2} + \frac{\zeta \mathrm{i}}{2} & \frac{1}{2} - \frac{\zeta}{2}
			\end{bmatrix}\!,\!
			\begin{bmatrix}
				\frac{1}{2} + \frac{\zeta}{2} & -\frac{\zeta}{2} + \frac{\zeta \mathrm{i}}{2} \\
				-\frac{\zeta}{2} - \frac{\zeta \mathrm{i}}{2} & \frac{1}{2} - \frac{\zeta}{2}
			\end{bmatrix}\!, \right. \nonumber \\ 
			&\ \ \left.
			\setlength{\arraycolsep}{2pt}
			\renewcommand{\arraystretch}{1.3}\!
			\begin{bmatrix}
				\frac{1}{2} - \frac{\zeta}{2} & -\frac{\zeta}{2} - \frac{\zeta \mathrm{i}}{2} \\
				-\frac{\zeta}{2} + \frac{\zeta \mathrm{i}}{2} & \frac{1}{2} + \frac{\zeta}{2}
			\end{bmatrix}\!,\!
			\begin{bmatrix}
				\frac{1}{2} - \frac{\zeta}{2} & \frac{\zeta}{2} + \frac{\zeta \mathrm{i}}{2} \\
				\frac{\zeta}{2} - \frac{\zeta \mathrm{i}}{2} & \frac{1}{2} + \frac{\zeta}{2}
			\end{bmatrix}\!
			\right\}\!,
		\end{align}
	\end{subequations}
	where the structure of $P_{\mathcal{A}}(a|\lambda_1,\lambda_2)$ is organized such that its rows correspond to measurement outcomes $a$, while its columns represent elements of the set $\Omega_{\mathrm{IN}}$ (as defined in Sec. \ref{sec3:level1}), e.g., the first column corresponds to $\lambda_1 = \lambda_2 = 1$, the second column corresponds to $\lambda_1 = 1$ and $\lambda_2 = 2$, and so forth. For any $\lambda_1, \lambda_2$, both $\sigma_{\lambda_1}^{\mathcal{B}_1}$ and $\sigma_{\lambda_2}^{\mathcal{B}_2}$ are Hermitian and positive semidefinite only when $\xi, \zeta \in [-{1}/{\sqrt{3}}, {1}/{\sqrt{3}}]$, in which $\xi$ and $\zeta$ are functions of $\nu$ and $\omega$, defined as follows
	\begin{subequations}\label{equ18}
		\begin{equation}\label{equ18-a}
			\xi(\nu,\omega) = \begin{cases} 
				\nu & \mathrm{ if }\ 0\leqslant \nu \leqslant\frac{1}{\sqrt{3}},\ 0\leqslant\omega\leqslant\nu,\\[2ex] 
				\sqrt{\nu^2\omega} & \mathrm{ if }\ \frac{1}{\sqrt{3}} \leqslant \nu \leqslant 1,\ 0\leqslant\omega\leqslant\frac{1}{3},\\[2ex]
				\sqrt{\nu\omega} & \mathrm{ if } \  \nu \geqslant \frac{1}{\sqrt{3}} ,\ \omega\geqslant\frac{1}{3},\ \nu\omega\leqslant\frac{1}{3},
			\end{cases}
		\end{equation}
		\begin{equation}\label{equ18-b}
			\zeta(\nu,\omega) = \begin{cases}
				\omega & \mathrm{ if }\ 0\leqslant \nu \leqslant\frac{1}{\sqrt{3}},\ 0\leqslant\omega\leqslant\nu,\\[2ex]
				\sqrt{\omega} & \mathrm{ if }\ \frac{1}{\sqrt{3}} \leqslant \nu \leqslant 1,\ 0\leqslant\omega\leqslant\frac{1}{3},\\[2ex]
				\sqrt{\nu\omega} & \mathrm{ if } \  \nu \geqslant \frac{1}{\sqrt{3}} ,\ \omega\geqslant\frac{1}{3},\ \nu\omega\leqslant\frac{1}{3}.
			\end{cases}
		\end{equation}
	\end{subequations}
	With this setup, the NLHS model can be systematically reconstructed. Particularly, within the region $\{{1}/{\sqrt{3}} \leqslant \nu \leqslant 1, 0 \leqslant \omega \leqslant {1}/{3}\}$, $\rho_1$ is entangled, whereas $\rho_2$ remains separable. Since $\mathcal{A}$ performs an entangled measurement, $\sigma_{\lambda_1}^{\mathcal{B}_1}$ is influenced by the states $\rho_1$ and $\rho_2$, while $\sigma_{\lambda_2}^{\mathcal{B}_2}$ remains independent of $\rho_1$.
	
	
	Although Eqs.~\eqref{equ17} and \eqref{equ18} are not the only forms that can be considered, our analysis establishes a significant result: for $\nu\omega > {1}/{3}$, it demonstrates the existence of network steering, i.e., $\mathcal{A}$ is capable of simultaneously steering the subsystems of both $\mathcal{B}_1$ and $\mathcal{B}_2$ through a single BSM.
		
\section{\label{sec5:level1}CONCLUSION}

	In this work, we have proposed a general definition of network steering scenarios, which also applies to standard steering scenarios. To measure steerability in quantum networks, we developed a neural network-based method. This method focused on identifying the closest network assemblage to a given one that admits the NLHS model, with the distance between these two assemblages serving as a quantitative measure of network quantum steerability. By casting this problem as a nonlinear optimization problem, we leveraged the causal information inherent in quantum networks and the principle of causality to solve it efficiently using neural networks. This process has facilitated efficient computation and yielded results consistent with established benchmarks, while also revealing novel findings.
	
	Upon application to standard steering scenarios, our method demonstrated high fidelity with results from Refs.~\cite{cavalcantiExperimentalCriteriaSteering2009,cavalcantiQuantitativeRelationsMeasurement2016,kuEinsteinPodolskyRosenSteeringIts2018,cavalcantiDetectionEntanglementAsymmetric2015}, thereby validating its accuracy and efficacy. Furthermore, we applied our method to measure steerability in the bilocal network steering scenario. Leveraging the numerical construction of an explicit NLHS model and the entanglement properties of network assemblages, we demonstrated that when the untrusted central party shares two-qubit isotropic states of different visibilities, $\nu$ and $\omega$, with two trusted endpoint parties and performs a single BSM, the network steering thresholds are determined by the curve $\nu \omega = {1}/{3}$.  
	
	This synergistic combination of ANNs and the NLHS model has offered a robust and scalable tool for measuring network quantum steerability. It provides a novel strategy for overcoming challenges in measuring quantum steerability in increasingly complex networks, positioning neural networks as a pivotal technology at the intersection of artificial intelligence and quantum information science.

\section*{Acknowledgment}
This work is supported by the National Natural Science Foundation of China (Grants No.62171056 and No.62220106012).

\appendix

\section{\label{app1:sec1}LOWER AND UPPER BOUNDS OF MULTIPARTITE QUANTUM STEERABILITY}
	
	In this section, we present a detailed derivation of the lower and upper bounds for $\mathcal{S}^{\mathrm{M}_1}$ and $\mathcal{S}^{\mathrm{M}_2}$. While our focus is primarily on the tripartite case, this approach can naturally extend to arbitrary n-party systems.
	
	For scenario 1-UNT, as detailed in Sec.~\ref{sec3:level1}, the distance between two assemblages, $\{\sigma_{a|x}^{\bar{\mathcal{B}}}\}_{a,x}$ and $\{ \sigma^{\mathrm{MLHS}_1}_{a|x}\}_{a,x}$, can be formulated as
	\begin{equation}\label{app:equ1}
		\begin{aligned}
			\mathcal{D}_{\mathrm{A}}(\{\sigma_{a|x}^{\bar{\mathcal{B}}}\}_{a,x},\{ \sigma^{\mathrm{MLHS}_1}_{a|x}\}_{a,x}) & = \frac{1}{N}\sum_{a,x} \mathcal{D}_{\mathrm{T}}(\sigma_{a|x}^{\bar{\mathcal{B}}}, \sigma^{\mathrm{MLHS}_1}_{a|x}), \\
			& = \frac{1}{2N} \sum_{a,x}\left \|\sigma_{a|x}^{\bar{\mathcal{B}}}-\sigma^{\mathrm{MLHS}_1}_{a|x}  \right \|_1,\\
			& \geqslant \frac{1}{2N} \sum_{a,x}\left \|\sigma_{a|x}^{\bar{\mathcal{B}}}-\sigma^{\mathrm{MLHS}_1}_{a|x}  \right \|_{\infty},\end{aligned}
	\end{equation}
	where the operator norm $\|\cdot\|_{\infty}$ is defined as: $\min \{\eta | -\eta \mathcal{I} \leqslant \cdot \leqslant \eta \mathcal{I} \}$. In accordance with this property, we establish the following lower bound for $\mathcal{S}^{\mathrm{M}_1}$
	\begin{equation}\label{app:equ2}
		\begin{aligned}\mathcal{S}^{\mathrm{M_1}}_{\mathrm{lb}}:= \min_{\sigma_{\lambda}^{\bar{\mathcal{B}}}}&\ \frac{1}{2N}\sum_{a,x} \left\|\sigma_{a|x}^{\bar{\mathcal{B}}}-\sigma^{\mathrm{MLHS_1}}_{a|x}\right\|_\infty,\\\mathrm{s.t.}  &\  \sigma^{\mathrm{MLHS_1}}_{a|x}=\sum_{\lambda}D(a|x,\lambda)\sigma_\lambda^{\bar{\mathcal{B}}} \ \  \forall a,x,\\&\  \sum_a \sigma^{\mathrm{MLHS_1}}_{a|x} =\rho^{\bar{\mathcal{B}}} \ \  \forall x, \\ & \ \sigma_{\lambda}^{\bar{\mathcal{B}}}\geqslant0,\ (\sigma_{\lambda}^{\bar{\mathcal{B}}})^{T_{\mathcal{B}_1}}\geqslant0\ \ \forall\lambda,\end{aligned}
	\end{equation}
	where $D(a|x, \lambda)$ represents the deterministic response function defined as $\delta_{a, \lambda(x)}$, and $T_{\mathcal{B}_1}$ denotes the partial transpose with respect to ${\mathcal{B}_1}$'s subsystem \cite{horodeckiSeparabilityMixedStates1996,cavalcantiDetectionEntanglementAsymmetric2015}. The above optimization problem can be reformulated as the following SDP
	\begin{equation}\label{app:equ3}
		\begin{aligned} \min_{\eta_{a,x},\sigma_{\lambda}^{\bar{\mathcal{B}}}}&\ \frac{1}{2N}\sum_{a,x}\eta_{a,x},\\\mathrm{s.t.} \ \ &\  -\eta_{a,x} \mathcal{I}\leqslant\sigma_{a|x}^{\bar{\mathcal{B}}}-\sum_{\lambda}D(a|x,\lambda)\sigma_{\lambda}^{\bar{\mathcal{B}}} \leqslant\eta_{a,x}\mathcal{I}\ \  \forall a,x,\\&\ \sum_{\lambda}\sigma_{\lambda}^{\bar{\mathcal{B}}}=\rho^{\bar{\mathcal{B}}} ,\\&\ \sigma_{\lambda}^{\bar{\mathcal{B}}}\geqslant0,\ (\sigma_{\lambda}^{\bar{\mathcal{B}}})^{T_{\mathcal{B}_1}}\geqslant0\ \ \forall\lambda.\end{aligned}
	\end{equation}
	
	To derive an upper bound for $\mathcal{S}^{\mathrm{M}_1}$, inspired by Ref.~\cite{cavalcantiQuantitativeRelationsMeasurement2016}, we introduce the concept of multipartite consistent steering robustness with 1-UNT, denoted as $\mathcal{S}_{\mathrm{MCSR_1}}$, which is defined as
	\begin{equation}\label{app:equ4}
		\begin{aligned}\min_{r,\pi_{a|x},\sigma_\lambda^{\bar{\mathcal{B}}}}&\  r,\\\mathrm{s.t.}\ \ &\  \frac{\sigma_{a|x}^{\bar{\mathcal{B}}}+r\pi_{a|x}}{1+r} = \sum_{\lambda}D(a|x,\lambda)\sigma_\lambda^{\bar{\mathcal{B}}} \ \  \forall a,x,\\&\ \sum_{a}\pi_{a|x} =\rho^{\bar{\mathcal{B}}} \ \  \forall x,\\
			&\ \sigma_{\lambda}^{\bar{\mathcal{B}}}\geqslant0,\ (\sigma_{\lambda}^{\bar{\mathcal{B}}})^{T_{\mathcal{B}_1}}\geqslant0\ \  \forall\lambda ,\\
			&\ \pi_{a|x}\geqslant0 \ \  \forall a,x,\\
			&\ r \geqslant0,\mathrm{Tr}\sum_{\lambda}\sigma_\lambda^{\bar{\mathcal{B}}}  =1.\end{aligned}
	\end{equation}
	This quantity quantifies the minimal level of arbitrary noise, denoted as $\{\pi_{a|x}\}_{a,x}$, required to transform an assemblage $\{\sigma_{a|x}^{\bar{\mathcal{B}}}\}_{a,x}$ into a ${\mathrm{MLHS_1}}$ assemblage. Importantly, the noise assemblage $\{\pi_{a|x}\}_{a,x}$ is constrained to preserve the same reduced state with the original assemblage, i.e., $\sum_a \pi_{a|x} = \sum_a \sigma_{a|x}^{\bar{\mathcal{B}}} = \rho^{\bar{\mathcal{B}}}$. Let $r^{\mathrm{opt}}$ and $\{\pi_{a|x}^{\mathrm{opt}}\}_{a,x}$ denote the optimal solutions obtained from the optimization problem in Eq.~\eqref{app:equ4}. Based on these solutions, we construct a new assemblage characterized by
	\[\sigma^{\mathrm{MCSR_1}}_{a|x}: = \frac{\sigma_{a|x}^{\bar{\mathcal{B}}}+r^{\mathrm{opt}}\pi_{a|x}^{\mathrm{opt}}}{1+r^{\mathrm{opt}}}.\]
	Following this construction, we establish the upper bound for $\mathcal{S}^{\mathrm{M}_1}$ as
	\begin{equation}\label{app:equ5}
		\mathcal{S}^{\mathrm{M_1}}_{\mathrm{ub}} : =\mathcal{D}_{\mathrm{A}}(\{\sigma_{a|x}^{\bar{\mathcal{B}}}\}_{a,x},\{ \sigma^{\mathrm{MCSR_1}}_{a|x}\}_{a,x}).
	\end{equation}
	Since $\mathcal{S}^{\mathrm{M_1}}_{\mathrm{ub}}$ is derived for a restricted noise, it is evident that $\mathcal{S}^{\mathrm{M}_1} \leqslant \mathcal{S}^{\mathrm{M_1}}_{\mathrm{ub}}$. The equality holds if and only if the assemblages $\{ \sigma^{\mathrm{MCSR_1}}_{a|x}\}_{a,x}$ and $\{ \sigma^{\mathrm{MLHS}_1}_{a|x}\}_{a,x}$ are equivalent.

		
	For the scenario of $2$-UNT, the lower bound of $\mathcal{S}^{\mathrm{M_2}}$ can be determined by solving
	\begin{equation}\label{app:equ6}
		\begin{aligned}\mathcal{S}^{\mathrm{M_2}}_{\mathrm{lb}}:= \min_{\sigma_{\lambda}^{\mathcal{B}}}&\ \frac{1}{2N_1N_2}\sum_{\bar{a},\bar{x}} \left\|\sigma_{\bar{a}|\bar{x}}^{\mathcal{B}}-\sigma^{\mathrm{MLHS_2}}_{\bar{a}|\bar{x}}\right\|_\infty,\\\mathrm{s.t.}&\  \sigma^{\mathrm{MLHS_2}}_{\bar{a}|\bar{x}}=\sum_{\lambda}D(\bar{a}|\bar{x},\lambda)\sigma_\lambda^{\mathcal{B}} \ \  \forall \bar{a},\bar{x},\\&\  \sum_{\bar{a}} \sigma^{\mathrm{MLHS_2}}_{\bar{a}|\bar{x}} =\rho^{\mathcal{B}} \ \  \forall \bar{x}, \\ & \ \sigma_{\lambda}^{\mathcal{B}}\geqslant0\ \ \forall\lambda,\end{aligned}
	\end{equation}
	where $D(\bar{a}|\bar{x},\lambda)=D(a_1|x_1,\lambda)D(a_2|x_2,\lambda)$ denotes the deterministic response function, given by $\delta_{a_1,\lambda{(x_1)}}\cdot\delta_{a_2,\lambda{(x_2)}}$. The above optimization problem is equivalent to the following formulation as a SDP
	\begin{equation}\label{app:equ7}
		\begin{aligned} \min_{\eta_{\bar{a},\bar{x}},\sigma_{\lambda}^{\mathcal{B}}}&\ \frac{1}{2N_1N_2}\sum_{\bar{a},\bar{x}}\eta_{\bar{a},\bar{x}},\\\mathrm{s.t.}\ \ &\  -\eta_{\bar{a},\bar{x}} \mathcal{I}\leqslant\sigma_{\bar{a}|\bar{x}}^{\mathcal{B}}-\sum_{\lambda}D(\bar{a}|\bar{x},\lambda)\sigma_{\lambda}^{\mathcal{B}}  \leqslant\eta_{\bar{a},\bar{x}}\mathcal{I}\ \  \forall \bar{a},\bar{x},\\&\ \sum_{\lambda}\sigma_{\lambda}^{\mathcal{B}}=\rho^{\mathcal{B}} ,\\&\  \sigma_{\lambda}^{\mathcal{B}}\geqslant0\ \ \forall\lambda.\end{aligned}
	\end{equation}

	Similarly, to derive an upper bound for $\mathcal{S}^{\mathrm{M}_2}$, we introduce the concept of multipartite consistent steering robustness with $2$-UNT , denoted as $\mathcal{S}_{\mathrm{MCSR_2}}$, which is defined as
	\begin{equation}\label{app:equ8}
		\begin{aligned}\min_{r,\pi_{\bar{a}|\bar{x}},\sigma_\lambda^{\mathcal{B}}}&\  r,\\\mathrm{s.t.}\ \  \ &\  \frac{\sigma_{\bar{a}|\bar{x}}^{\mathcal{B}}+r\pi_{\bar{a}|\bar{x}}}{1+r} = \sum_{\lambda}D(\bar{a}|\bar{x},\lambda)\sigma_\lambda^{\mathcal{B}} \ \  \forall \bar{a},\bar{x},\\
			&\ \sum_{\bar{a}}\pi_{\bar{a}|\bar{x}} =\rho^{\mathcal{B}} \ \  \forall \bar{x},\\
			&\ \sigma_{\lambda}^{\mathcal{B}}\geqslant0\ \  \forall\lambda ,\\
			&\ \pi_{\bar{a}|\bar{x}}\geqslant0 \ \  \forall \bar{a},\bar{x},\\
			&\ r \geqslant0,  \text{Tr}\sum_{\lambda}\sigma_\lambda^{\mathcal{B}}  =1.\end{aligned}
	\end{equation}
	Let $r^{\mathrm{opt}}$ and $\{\pi_{\bar{a}|\bar{x}}^\mathrm{opt}\}_{\bar{a},\bar{x}}$ denote the optimal solutions to Eq.~\eqref{app:equ8}, and then we construct a new assemblage
	\[
	\sigma^{\mathrm{MCSR_2}}_{\bar{a}|\bar{x}}: = \frac{\sigma_{\bar{a}|\bar{x}}^{\mathcal{B}}+r^{\mathrm{opt}}\pi_{\bar{a}|\bar{x}}^{\mathrm{opt}}}{1+r^{\mathrm{opt}}}.
	\]
	The upper bound for $\mathcal{S}^{\mathrm{M_2}}$ is thereby established as
	\begin{equation}\label{appb:equ13}
		\mathcal{S}^{\mathrm{M_2}}_{\mathrm{ub}} : =\mathcal{D}_{\mathrm{A}}(\{\sigma_{\bar{a}|\bar{x}}^{\mathcal B}\}_{\bar{a},\bar{x}},\{ \sigma^{\mathrm{MCSR_2}}_{\bar{a}|\bar{x}}\}_{\bar{a},\bar{x}}).
	\end{equation}
	Since Eqs.~\eqref{app:equ3}, \eqref{app:equ4}, \eqref{app:equ7}, and \eqref{app:equ8} all correspond to SDP problems, they can be efficiently solved using state-of-the-art convex optimization toolkits, such as CVX \cite{cvx} for MATLAB.

\end{document}